\documentclass[aip,reprint,superscriptaddress]{revtex4-2}
\usepackage{upgreek}
\usepackage{graphicx}
\usepackage{bm}
\usepackage{amsmath, amsthm, amssymb}
\usepackage{latexsym}
\usepackage{dcolumn}

\usepackage{mathtools}

\DeclarePairedDelimiterX\braket[2]{\langle}{\rangle}{#1 \delimsize\vert #2}
\begin{document}
	
	\preprint{APS/123-QED}
	
	\title{Optical dynamic nuclear polarization of $^{13}$C spins in diamond at a low field with multi-tone microwave irradiation}

\author{Vladimir Vladimirovish \surname{Kavtanyuk}}
\affiliation{Quantum Magnetic Imaging Team, Korea Research Institute of Standards and Science, Daejeon 34113, Republic of Korea}
\author{Hyun Joon \surname{Lee}}
\affiliation{Radio \& Satellite Research Division, Electronics and Telecommunications Research Institute, Daejeon 34129, Republic of Korea}
\author{Sangwon \surname{Oh}}
\affiliation{Quantum Magnetic Imaging Team, Korea Research Institute of Standards and Science, Daejeon 34113, Republic of Korea}
\author{Keunhong \surname{Jeong}}
\affiliation{Korea Military Academy, Seoul 01805, Republic of Korea}
\author{Jeong Hyun  \surname{Shim}}
\email{jhshim@kriss.re.kr}
\affiliation{Quantum Magnetic Imaging Team, Korea Research Institute of Standards and Science, Daejeon 34113, Republic of Korea}
\affiliation{Department of Medical Physics, University of Science and Technology, Daejeon 34113, Republic of Korea}
	
\date{\today}
	
	\begin{abstract}
	Most of dynamic nuclear polarization (DNP) has been requiring helium cryogenics and strong magnetic fields for a high degree of polarization. In this work, we instead demonstrate an optical hyperpolarization of naturally abundant $^{13}$C nuclei in a diamond crystal at a low magnetic field and an ambient temperature. It exploits continuous irradiations of pump laser for polarizing electron spins of nitrogen vacancy centers and microwave for transferring the induced polarization to $^{13}$C nuclear spins. Triplet structures corresponding to $^{14}$N hyperfine splitting were clearly observed in the spectrum of $^{13}$C polarization. The powers of microwave irradiation and pump laser were optimized. By simultaneously irradiating three microwave frequencies matching to the peaks of the triplet, we achieved a $^{13}$C bulk polarization of 0.113 $\%$, leading to an enhancement of about a factor of 90,000 over the thermal polarization at 17.6 mT. We believe that the multi-tone irradiation can be universally adopted to further enhance the $^{13}$C polarization at a low magnetic field.
	\end{abstract}
	
	\maketitle
	
	
	\section{\label{sec:level1}INTRODUCTION}
\begin{figure}[t!]
		\includegraphics[width = 8.5 cm]{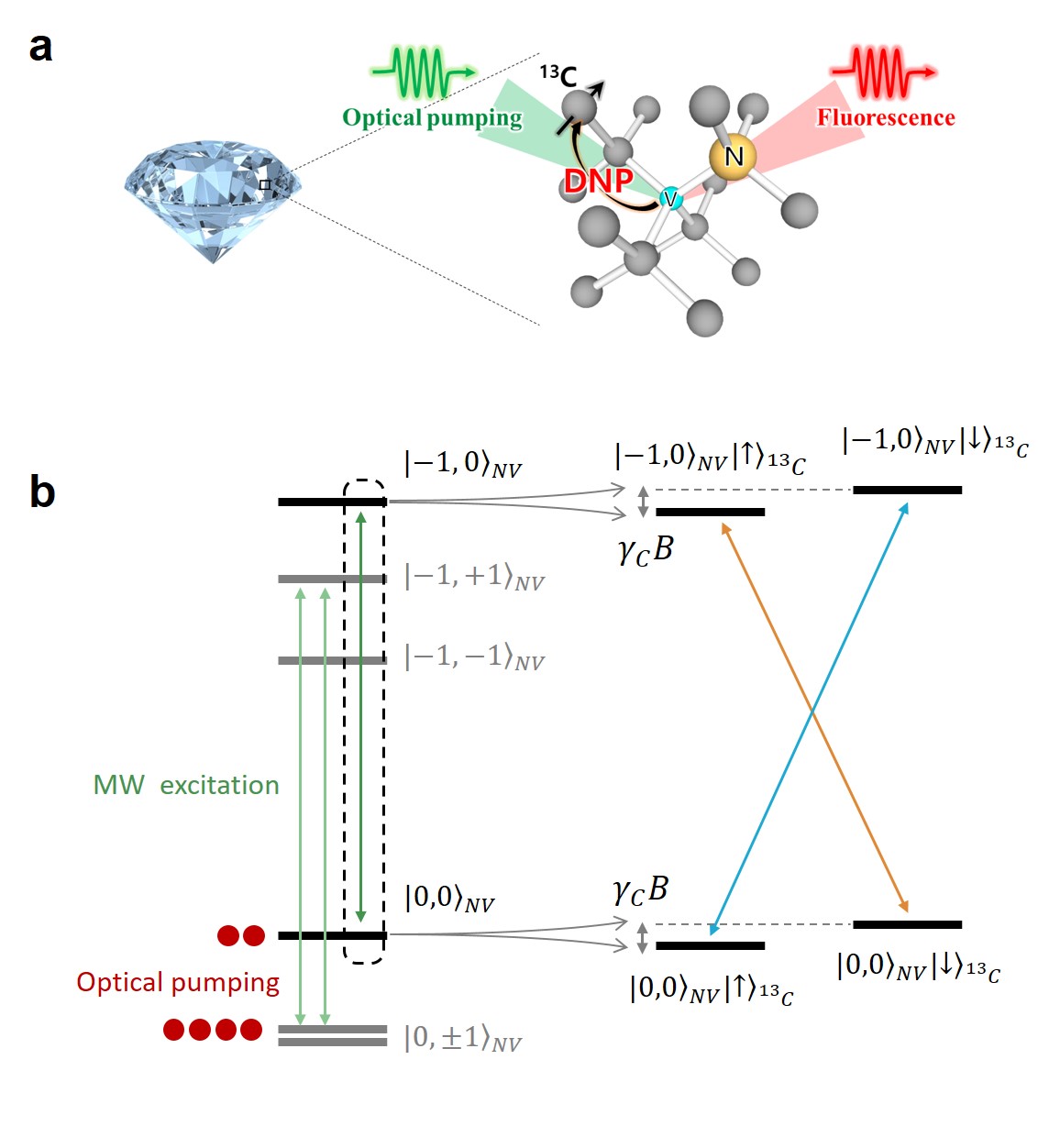}
		\caption{\label{fig:fig1} (a) The schematic illustration of optical DNP process in diamond. The optically induced polarization in NV spins will be transferred to bulk $^{13}$C nuclear spins. (b) The part of energy level structure associated with NV, $^{14}$N, and distant $^{13}$C spins is given. At a low field, $^{13}$C nuclear Zeeman energy $\gamma_cB$ is weaker than the $^{14}$N hyperfine splitting (2.16 MHz) in NV spins. The arrows indicates the possible transitions induced by MW irradiations used in optical DNP.}
	\end{figure}

	Dynamic nuclear polarization (DNP) has been a technological breakthrough, which can significantly boost the signal-to-noise ratio in nuclear magnetic resonance (NMR) and magnetic resonance imaging (MRI) applications \cite{ArdenkjaerLarsen2003, Joo2006, Thankamony2017, Rankin2019, Bucher2020, Krummenacker2012, Denysenkov2017}. Enhanced signal enables extracting substantial amount of information at molecular levels for a wide-range of chemical, biological and physical processes \cite{Golman2006, Barnes2008, Rossini2012, Puebla2013, Masion2017, Flori2018, Morze2019, Park2021,  Abhyankar2021}. Recently, optical dynamic nuclear polarization, in which electronic polarization of negatively-charged nitrogen-vacancy (NV) centers is transferred to bulk $^{13}$C nuclear spins (Fig. \ref{fig:fig1}(a)), has been demonstrated \cite{Fischer2013, Wang2013, Alvarez2015, King2015, Ajoy2018, Ajoy2018a,  Ajoy2020, Ajoy2021, Parker2019}. Besides continuous-wave and simultaneous irradiations of microwave and pump laser, a variety of NV-based optical DNP techniques have been introduced including pulsed irradiation of microwave (MW) fields satisfying Hartman-Hahn condition \cite{Scheuer2016}, the cross polarization at the level anti-crossing in the ground state \cite{Fischer2013} and the excited state \cite{Wang2013}. Frequency sweeping \cite{Ajoy2018, Ajoy2018a, Ajoy2020, Ajoy2021} was exploited to hyperpolarize diamond powders or nanodiamonds that can potentially be used for molecule-targeted in-vivo imaging with the advantage of a long spin life time \cite{Cassidy2013, Rej2015, Kwiatkowski2017, Hu2018, Seo2018}.

	Owing to the optically induced polarization, NV-based optical DNP does not need cryogenic apparatus and can operate in low magnetic fields of tens of millitesla. The magnetic fields in this range are easily accessible in experiments, however certainly not favorable for efficient hyperpolarization. Firstly, both a weak thermal nuclear polarization and a short T$_1$ relaxation time are disadvantageous. In addition, the positive and negative nuclear polarization spectra, being the hallmark of solid-state dynamic nuclear polarization, are separated by the order of nuclear Zeeman splitting\cite{Wenckenbach2016}. Thus, in low fields the partial overlap of positive and negative nuclear polarizations may occur, and this will eventually reduce the net nuclear polarization. Nevertheless, such nuclear polarization spectrum, which reflects the weak $^{13}$C Zeeman splitting in low magnetic fields, has not been observed yet in optical DNP studies of diamonds. If $^{13}$C resonance frequency is less than $^{14}$N hyperfine splitting (2.16 MHz) of NV spin (Fig. \ref{fig:fig1}(b)), the nuclear polarization spectrum would be duplicated three times. Not surprisingly such triplet structure has not been reported either.
	
	In the present study, we conducted NV-based optical DNP of bulk $^{13}$C nuclear spins in diamond at a low field and an ambient temperature. Applied field was 17.6 mT, which corresponds to $^{13}$C resonance frequency of approximately 0.2 MHz. MW and laser were continuously irradiated, and their optimal power levels were characterized. In $^{13}$C nuclear polarization spectrum, we clearly observed the triplet structure showing the $^{14}$N hyperfine splitting. As anticipated, nitrogen nuclear spin does not participate in the polarization transfer process except for in determining the resonance frequency of NV spins through the hyperfine interaction. By exploiting such independence, we irradiated triple microwave frequencies simultaneously for further boosting, and achieved a $^{13}$C nuclear polarization of 0.113 $\%$, which is about 1.7 times higher compared to DNP by single MW frequency irradiation.

\section{Experimental methods}	
We used a HPHT-grown diamond crystal with naturally abundant $^{13}$C nuclear spins. NV centers with a concentration of 1.25 ppm are fabricated via electron irradiation and thermal annealing process. We developed an optical DNP system, which includes low-field region for optical pumping, high-field region (6 T) for $^{13}$C NMR, and a rapid shuttling device for displacing the hyperpolarized diamond crystal to the high-field region in 2 s. From  $^{13}$C NMR signals at 6 T, enhancement factors and absolute polarization levels were estimated. $^{13}$C NMR and the timing controls were conducted with a commercial NMR console. (For more detailed explanations, see the appendix section.)

	\section{\label{sec:level2} Experimental Results}

	\subsection{\label{sec:level1}$^{13}$C polarization spectrum}
	
\begin{figure}[b]
		\includegraphics[width = 6.5 cm]{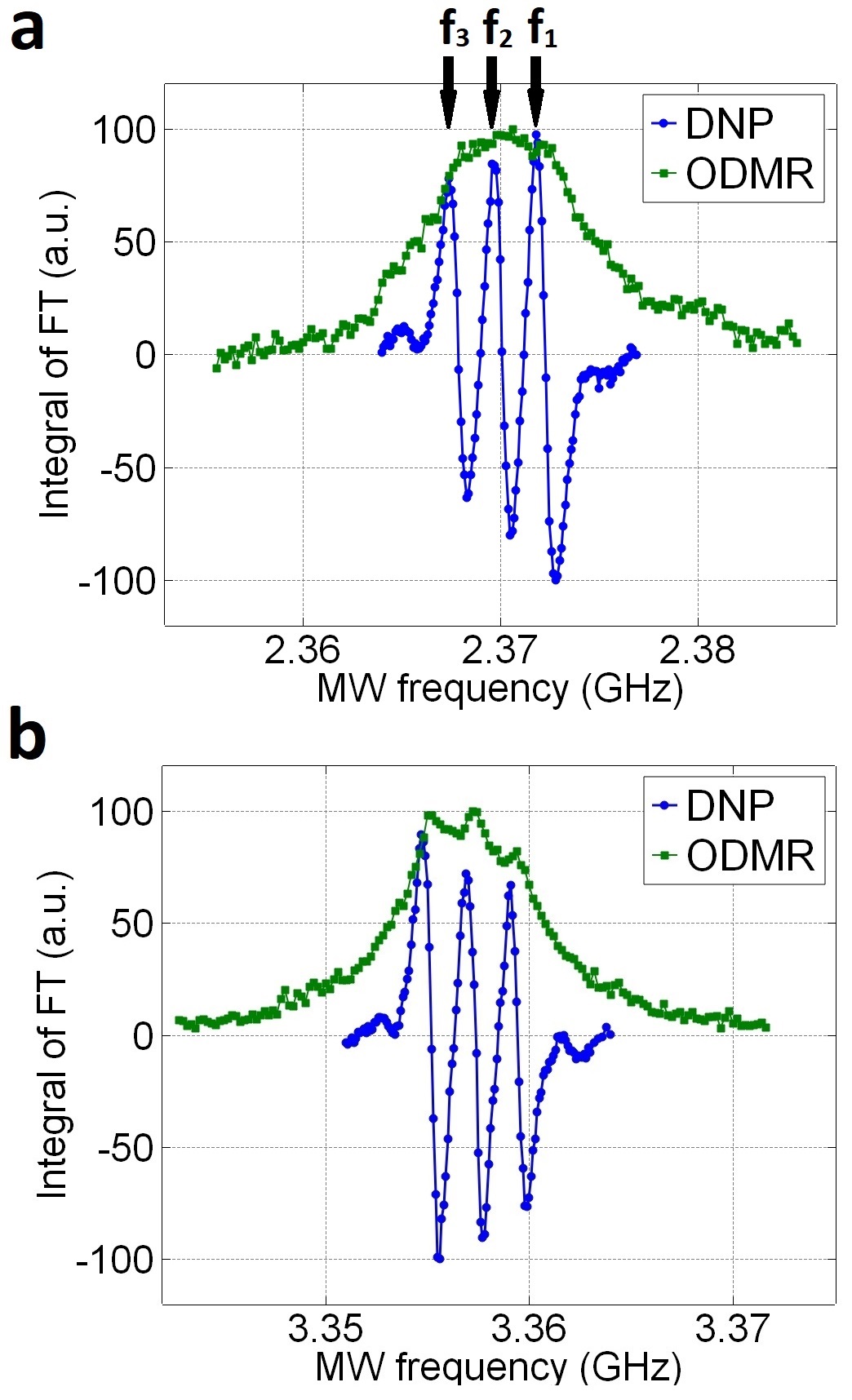}
		\caption{\label{fig:fig2} Normalized values of integrated Furrier Transform (FT) signals of the real part plotted as a function of the MW frequency (blue line) and ODMR signal with normalized intensity (green line) at 17.6 mT. Each data point for each MW frequency was measured with the same experimental conditions. (a) -1 state and  (b) +1 state of NV spins.}
	\end{figure}

	We first investigated $^{13}$C polarization spectrum. Varying the frequency of MW irradiated for optical DNP, we recorded the $^{13}$C NMR signal as represented by the blue lines in Fig. \ref{fig:fig2}. The green curves illustrate optically-detected magnetic resonance (ODMR) lines of NV center, including the transitions of m$_s$ = 0 $\rightarrow$ m$_s$ = -1 (Fig. \ref{fig:fig2}(a)) and m$_s$ = 0 $\rightarrow$ m$_s$ = +1 (Fig. \ref{fig:fig2}(b)). All these measurements were performed through NV spins of $\langle$111$\rangle$ orientation, which is parallel with the direction of magnetic field. Each data point was averaged by 10 measurements.  The recording sequence of $^{13}$C NMR started with saturating $^{13}$C spins with a series of 90 pulses, which destroyed a residual $^{13}$C polarization and ensured non-polarization. Triplet structures were clearly observed for both NV spin transitions. The $^{14}$N (I = 1) of the NV center splits the NV spin levels each into three hyperfine sublevels with an energy splitting of 2.16 MHz, and this explains the triplet shown in Fig. \ref{fig:fig2}. Although ODMR does not exhibit the $^{14}$N hyperfine splitting, it is visible in the $^{13}$C polarization spectra. Notably, the signs of the $^{13}$C polarization are identical for both NV spin transitions.
	
	The results in Fig. \ref{fig:fig2} are obtained at a magnetic field of 17.6 mT with a diamond containing natural abundance of $^{13}$C nuclear spins. A similar measurement was performed in Ref.\cite{Alvarez2015}, where a  magnetic field of 18 mT was applied. However, the $^{13}$C polarization spectrum showed no triplet structure. Since a diamond of 10$\%$ enriched $^{13}$C was used in Ref.\cite{Alvarez2015}, strong hyperfine interactions between NV and proximate $^{13}$C spins may dominate over the hyperfine interaction with the $^{14}$N spin. In Ref.\cite{Parker2019}, a higher magnetic field of 473 mT was applied, corresponding to a $^{13}$C resonance frequency of 5 MHz. In this case, the overlap of three spectra with a separation of 2.16 MHz wipes the triplet structure, resulting in a single curve.

\begin{figure}[b]
		\includegraphics[width = 7.5 cm]{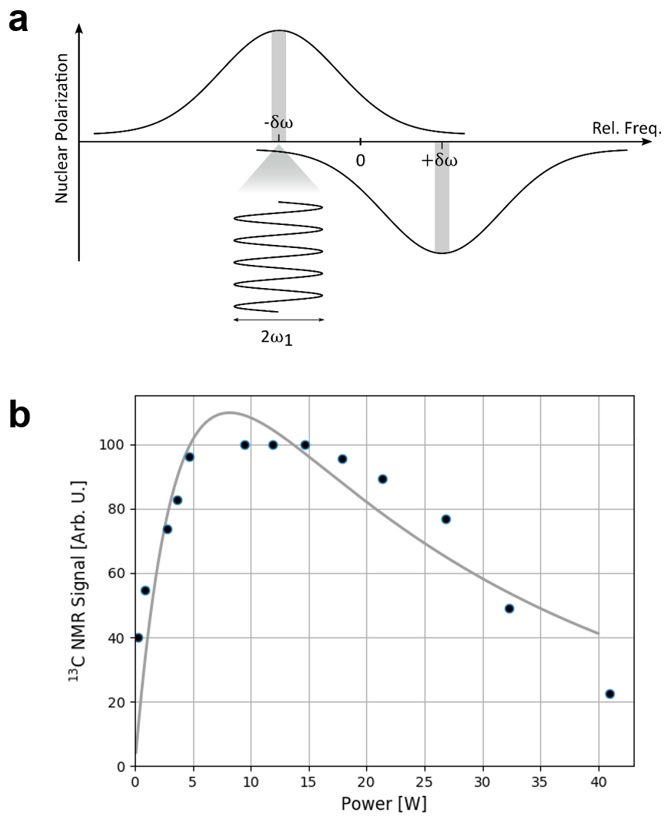}
		\caption{\label{fig:fig3} (a) Illustration of MW irradiation influences on nuclear polarization of $^{13}$C. (b) Normalized integrals of FT signals from hyperpolarized $^{13}$C vs MW power.}
	\end{figure}
	
	\subsection{\label{sec:level2} Optimal  microwave power}
	
	To improve the $^{13}$C polarization, we tried to optimize experimental parameters, such as MW and laser powers. Particularly at a low magnetic field, a strong MW field may not be beneficial. As the separation between positive and negative peaks in the nuclear polarization spectrum is proportional to the nuclear Zeeman splitting \cite{Wenckenbach2016}, the strong MW that is on-resonant with one polarization can also induce the transition of the opposite polarization, which is off-resonant as illustrated in Fig. \ref{fig:fig3} (a). Then, the opposite polarization will be imposed and eventually lead to a lower net nuclear polarization. Figure \ref{fig:fig3} (b) shows the $^{13}$C nuclear polarization as a function of MW power. The polarization rises initially but decays afterwards as the MW power increases. From the curve in Fig. \ref{fig:fig3}, we can determine the optimal MW power, near 10 W. The solid line is the fitted curve with two exponents for a guide to eyes. A similar result has been reported in Ref.\cite{Alvarez2015}, in which it is explained by a transition from selective regime to $\Lambda$ regime due to stronger MW fields.
		
	\subsection{\label{sec:level3} Optimal laser power density}
	
\begin{figure}[t]
		\includegraphics[width = 9 cm]{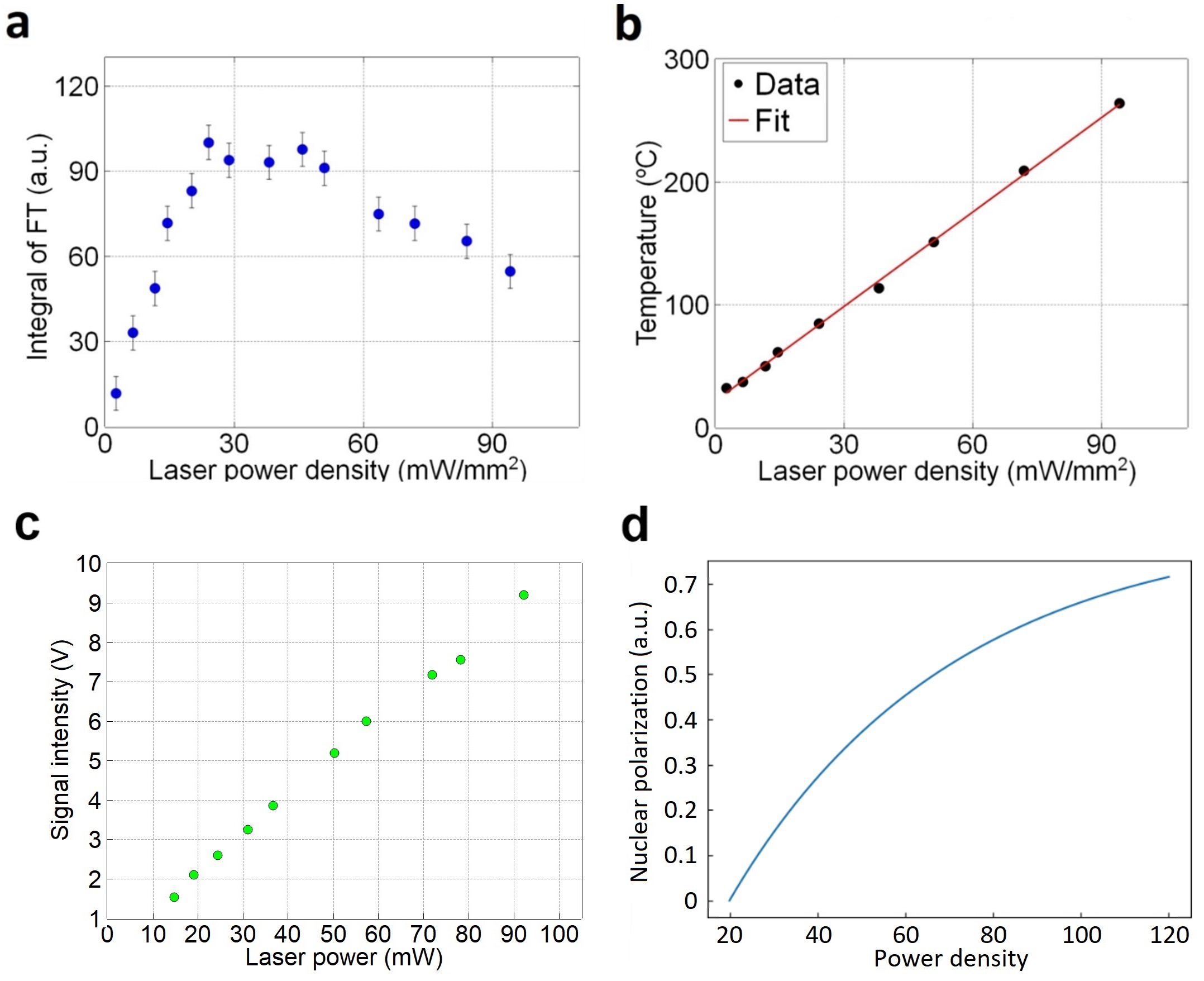}
		\caption{\label{fig:fig4}(a) Normalized integrals of FT signals from hyperpolarized $^{13}$C vs laser power density. (b) Laser power vs diamond temperature. (c) Laser power vs intensity of diamond's photoluminescence. (d) The curve of the $^{13}$C polarization expected by using the results in Fig. \ref{fig:fig4} (b) and (c). }
	\end{figure}

	We investigated how the $^{13}$C polarization varies with the pump laser power. Figure \ref{fig:fig4} (a) shows that the maximum $^{13}$C polarization signal was measured at a laser power density of 30 mW/mm$^2$. After the maximum, the polarization was gradually reduced as laser power increased. Initially, we speculated that the $^{13}$C polarization decrease due to the rise of diamond temperature. The temperature of the diamond crystal was measured from the position of zero-field splitting. From the ODMR spectrum, the peak positions of the two transitions, m$_s$ = 0 $\rightarrow$ m$_s$ = -1 and m$_s$ = 0 $\rightarrow$ m$_s$ = +1, can be obtained. Then, their mean results in the zero-field splitting with the conversion factor, $\frac{dD}{dT} \cong - 74$ kHz/K\cite{Acosta2010}. As shown in Fig. \ref{fig:fig4} (b), the temperature rise induced by the pump laser was linearly proportional to the laser power density. At the optimal near 30 mW/mm$^2$, the diamond temperature was raised more than 100 K. Temperature rise will reduce the thermal polarization of $^{13}$C nuclear spins. One may attribute the decrease of the $^{13}$C polarization to such laser-induced heating.
	
	We found out, however, that the laser-induced heating is not enough to explain the $^{13}$C polarization decrease. It's because the NV electronic spin polarization increases with the laser power. Figure \ref{fig:fig4} (c) shows the variation of the fluorescence intensity from NV spins in the diamond. Apparently, the NV spin polarization was far from the saturation and still increasing linearly. The enhancement factor is proportional to the NV spin polarization ($P_{\mathrm{NV}}$). Then, $^{13}$C polarization ($P_{\mathrm{Hyper}}$) obtained by the optical DNP becomes proportional to the NV spin polarization ($P_{\mathrm{NV}}$) times the $^{13}$C thermal polarization ($P_{\mathrm{Thermal}}$) under the pump laser. Given a pump laser density $\sigma_{\mathrm{Laser}}$, $P_{\mathrm{NV}}$ and $P_{\mathrm{Thermal}}$ can be expressed as below,

\begin{eqnarray}
&&P_{\mathrm{NV}} = \beta  \sigma_{\mathrm{Laser}}, \nonumber \\
 &&P_{\mathrm{Thermal}}= \frac{c}{300+\alpha \sigma_{\mathrm{Laser}}}
\end{eqnarray}
, in which $\alpha$ and $\beta$ are the slopes of the curves in Fig. \ref{fig:fig4} (b) and (c), respectively. ($c$ is a constant.) The resulting curve of $P_{\mathrm{Hyper}}$  is shown in Fig. \ref{fig:fig4} (d), which is certainly inconsistent with the result in Fig.  \ref{fig:fig4} (a).

	\subsection{\label{sec:level3} $^{13}$C hyperpolarization by multi-tone MW frequencies}
	
	Since multiple peaks were observed in the $^{13}$C polarization spectrum (Fig. \ref{fig:fig2}), we can enhance further the $^{13}$C polarization by simultaneous irradiation of multiple MW frequencies corresponding to those peaks. To estimate the $^{13}$C polarization obtained through the optical DNP, we compared the $^{13}$C NMR spectra with that of thermal polarization at 6 T (300 K). The thermal polarization of $^{13}$C in the diamond was measured for 4 days, with 400 averages and 20 min between each scan. The hyperpolarization of $^{13}$C in the diamond was performed by optical DNP using the transition from m$_s$ = 0 to m$_s$ = -1 shown in Fig. \ref{fig:fig2} (a). The diamond was irradiated for 120 s by the pump laser beam with a power density of 25 mW/mm$^2$ and by the MW field with a power of 11 W. Due to the laser irradiation the diamond was heated up to about 400 K.

\begin{figure}[t]
		\includegraphics[width = 7.5 cm]{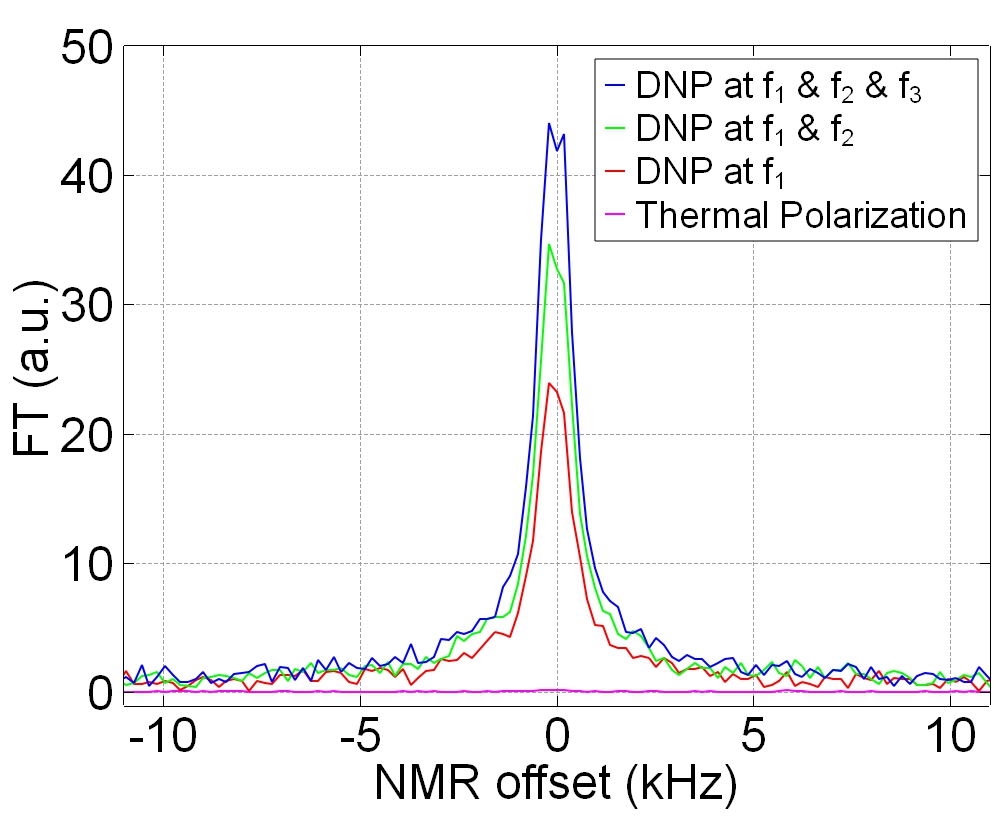}
		\caption{\label{fig:fig5} Comparison between thermally polarized $^{13}$C at 6 T and hyperpolarized $^{13}$C nuclei obtained by DNP method for -1 state of aligned NV$^-$ at 17.6 mT. All the measurements were performed on the same diamond at the room temperature. The pink line represents FT signal obtained from thermally polarized $^{13}$C nuclei with 400 averages. The red line represents FT signal obtained by MW excitation at the first hyperfine splitting corresponding to 0.068$\%$ of $^{13}$C hyperpolarization. The green line represents FT signal obtained by MW excitation at the first and the second hyperfine splitting corresponding to 0.094$\%$ of $^{13}$C hyperpolarization. The blue line represents FT signal obtained by MW excitation at the first, the second and the third hyperfine splitting corresponding to 0.113$\%$ of $^{13}$C hyperpolarization. The combined polarization was expected to be higher, we are not sure the reason of the lower polarization value.}
	\end{figure}

	Figure \ref{fig:fig5} shows the results of the multiple MW irradiation. The number of the MW frequencies have been increased from one (f$_1$) to three (f$_1$, f$_2$, f$_3$). The three MW frequencies are indicated by the arrows in Fig. \ref{fig:fig2} (a). Clearly, the $^{13}$C polarization increases with the number of MW frequency. It was 0.068 \% with f$_1$ only, while 0.113 \% with all three frequencies. The enhancement ratio is 1.7, which is less than the ideal value of about 2.5 predicted by the summed intensities in the Fig. \ref{fig:fig2} (a). This may be due to the effect of off-resonant excitations described above. The polarization of 0.113 \% corresponds to the enhancement of 90,000 times over the in situ thermal polarization at 17.6 mT and the diamond temperature as the calculation shown in the appendix.
	
	\section{\label{sec:level3}Conclusion}
	
	We have demonstrated the optical DNP results obtained with a diamond crystal of natural abundance of $^{13}$C nuclei at room temperature and a low magnetic field of 17.6 mT.  We observed a triplet structure in $^{13}$C nuclear polarization spectrum, in which the splitting of 2.16 MHz attributes to $^{14}$N hyperfine interaction in NV spin. The bulk $^{13}$C nuclear spin polarization reached 0.113\%. To obtain such $^{13}$C hyperpolarization, we have found optimal values of laser power density and MW power. In addition, we adopted multi-tone MW excitation. Simultaneous irradiation of three MW frequencies, matched to the peak frequencies in the $^{13}$C nuclear spectrum, results in an improvement of 1.7 times. We believe this multi-tone irradiation scheme can be universally exploited to further enhance the $^{13}$C polarization at a low magnetic field.
		
		\begin{acknowledgments}
		This work was supported by a grant (GP2021-0010) from Korea Research Institute of Standards and Science, Institute of Information and communications Technology Planning $\&$ Evaluation (IITP) grants funded by the Korea government (MSIT) (No.2019-000296, No.2019-0-00007), the Ministry of Science and ICT in Korea via KBSI (Grant No. C123000).
		\end{acknowledgments}
		
		\appendix
		
		\section{Diamond sample}
		
		All the experimental data, provided in this paper, have been measured from only one industrial diamond of 42 mg. It was synthesized by the high-pressure high-temperature method and was electron irradiated at 1 MeV followed by annealing at about 800 $^o$C. It contains about 50 ppm of P$_1$ centers, 1.25 ppm of NV centers and the natural abundance of $^{13}$C. The spin concentrations in the diamond were estimated by performing electron spin resonance (ESR). The peak intensities in ESR spectra were compared to that from a diamond crystal having a known P1 concentration. The thickness of the diamond is about 1 mm and its form can be considered as the equilateral triangle with side equal about 4 mm. The diamond is $\langle$111$\rangle$ surface-oriented, meaning that if a magnetic field is aligned along one of the NV center orientations, the other three orientations would have the angle of 109.5$^o$ away from the field direction.
		
			\section{Estimation of $^{13}$C nuclear hyperpolarization and enhancement factor}
		
		The hyperpolarization of $^{13}$C in the diamond obtained at $B_{\mathrm{EM}}$ = 17.6 mT has been estimated based on the thermal polarization of the same diamond obtained at $B_{SM}$ = 6 T. So, P$_{\mathrm{Hyper}}$ - the hyperpolarization of $^{13}$C has been calculated by the equation:
		
		\begin{equation}
			P_{\mathrm{Hyper}} = \frac{S_{\mathrm{Hyper}}}{S_{\mathrm{Thermal}}} * P_{\mathrm{Thermal}}
		\end{equation}
		
		where $S_{\mathrm{Hyper}}$ is the integral of FT signal measured from a hyperpolarized $^{13}$C nuclei in the diamond, $S_{\mathrm{Thermal}}$ is the integral of FT signal measured from thermally polarized $^{13}$C nuclei in the diamond and $P_{Thermal}$ = $\hbar$$\gamma$$B_0/2kT_R$ is the value of thermal polarization at the room temperature ($T_R$ = 297 K).
		
		The enhancement factor ($\varepsilon$) of $^{13}$C hyperpolarization is about 90,000 over thermal equilibrium. It has been calculated by this equation:
		
		\begin{equation}
		\varepsilon = \frac{S_{\mathrm{Hyper}}}{S_{\mathrm{Thermal}}} * \frac{B_{\mathrm{SM}}}{B_{\mathrm{EM}}} * \frac{T_{L}}{T_{R}}
		\end{equation}
	
		where $T_L$ is diamond temperature under laser irradiation during DNP process (Fig. \ref{fig:fig4}, b).
		
		\section{Experimental Setup}
		
\begin{figure}
		\includegraphics[width = \columnwidth]{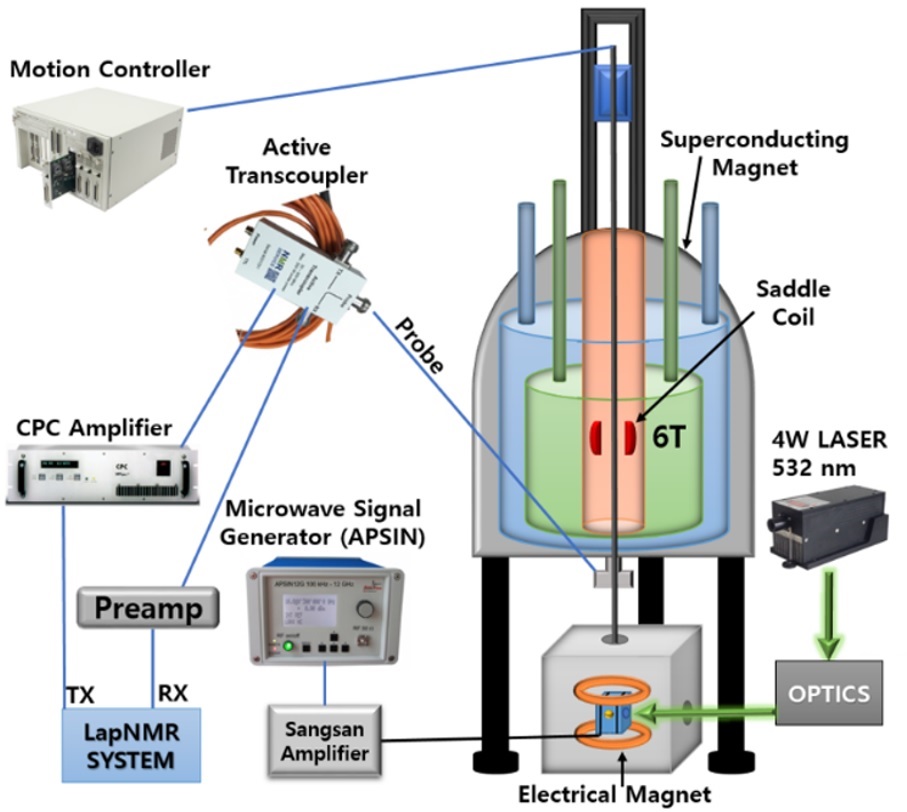}
		\caption{\label{fig:fullsetup}Full experimental setup. The right part shows schematic of our entire experimental setup. The left part shows small items of the setup in details. }
		\end{figure}

		Our experimental setup is sophisticated system capable of hyperpolarizing $^{13}$C in diamonds using DNP method and measuring polarization using NMR method. This section will thoroughly describe the experimental setup (Fig. \ref{fig:fullsetup}) and a purpose of each hardware.
		
		\subsection{\label{app:subsec}Superconducting magnet and NMR hardware}
		
		The NMR signals, presented in this paper, have been measured at the magnetic field of a superconducting magnet from OXFORD. It has a diameter of 70 cm and a height of 97 cm. The magnet produces a homogeneous magnetic field of about 6 T in its central region. There is a cylindrical cavity with a diameter of about 5.5 cm in the center of the magnet through its entire height. A handmade saddle coil is installed inside of the cavity in the homogeneous region for measuring polarization from samples. The coil is connected to a NMR probe (Fig. \ref{fig:fullsetup}), which is a LC circuit made of parallel and serial connected variable capacitors. By manipulating the capacitance, we can match the impendence and set a required resonance frequency for the saddle coil ($\sim$64.237 MHz for $^{13}$C). The probe is subsequently connected to an active-transcoupler and to a commercial spectrometer LapNMR from Tecmag. The purpose of the transcoupler is to manage transmitted signals from the spectrometer and received NMR signals from the saddle coil. The spectrometer is controlled by TNMR software, which allows us to set all required parameters for generating proper RF signals and acquire NMR signals. Transmitted RF signals from LapNMR is increased by CPC 9T100M power amplifier. NMR signals captured by the saddle coil is increased by two +35 dB preamplifiers from NMR service.
		
		\subsection{\label{app:subsec} MW hardware}
		
		One of the aspect in successful hyperpolarization of $^{13}$C in a diamond is application of MW field with enough power. In our experiment, MW signals are produced by APSIN12G and amplified by Sungsan TS2458BBP0 with a maximum power output of 50 W. A handmade Helmholtz coil with a diameter of 13 mm is used to generate MW fields (Fig. \ref{fig:fullsetup}). Due to impedance mismatch between the coil and other 50 Ohm hardware, most of applied power to the coil gets reflected. So, an actual power which can pass through the MW coil is less than 1$\%$ from the applied power. For this reason PE83CR1004 circulator with a 50 W termination is used to protect the amplifier from the high reflection power. The MW coil is installed inside of an electrical Helmholtz magnet (Fig. \ref{fig:fullsetup}) powered by a high precision DC power supply 2280S-32-6. The electrical magnet is installed right below the superconducting magnet and used for DNP measurements, since it provides a good homogeneity and the easy way to vary the magnetic field.
		
		\subsection{\label{app:subsec}Laser with optics}
		
\begin{figure}[b]
		\includegraphics[width = 8.5 cm]{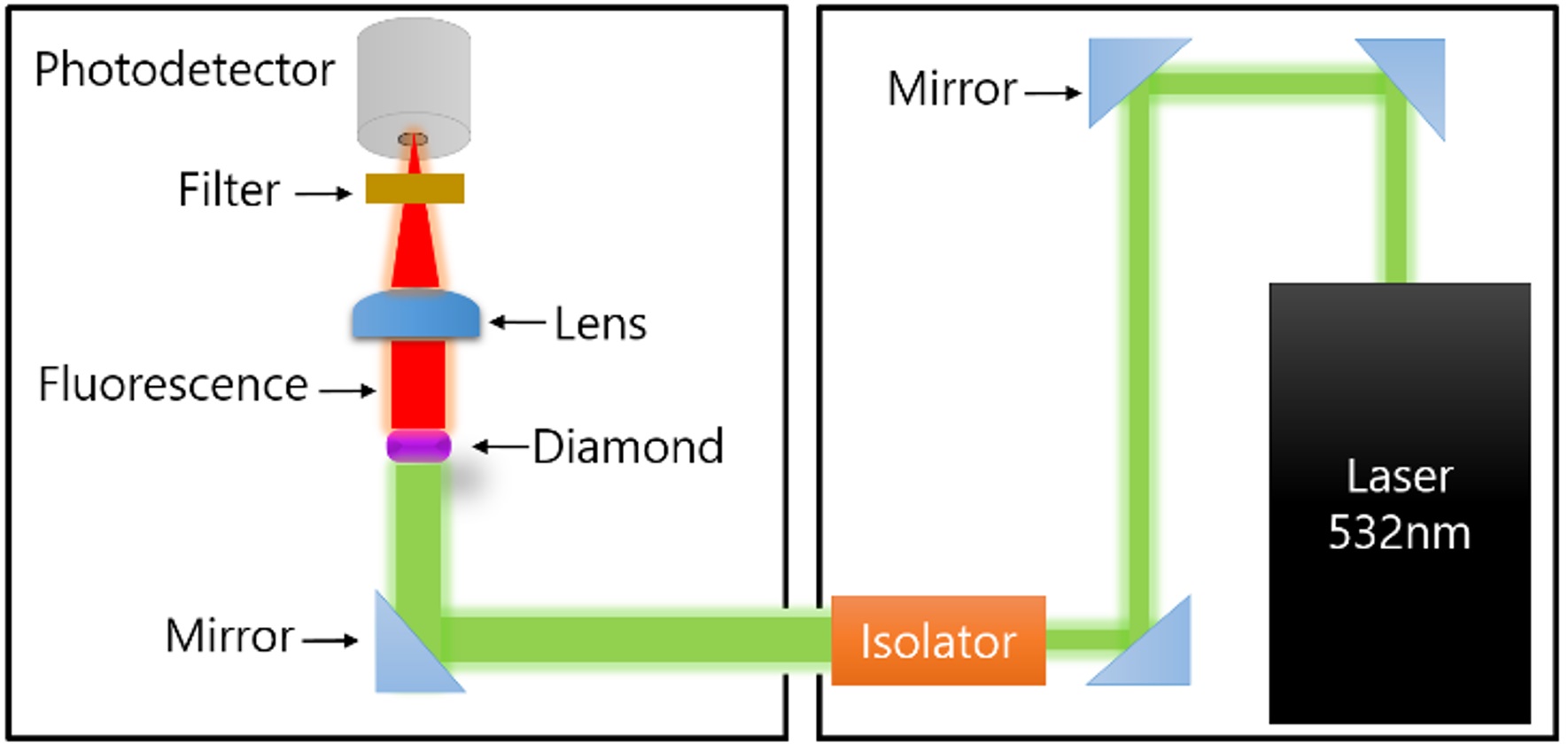}
		\caption{\label{fig:laser} Schematic of the laser optical setup.}
		\end{figure}

		Our laser produces vertically polarized 532 nm green light with maximum power output of 4 W and beam diameter of 2 mm. The delivery of the beam light from the laser to the diamond is performed by two separated optical setups (Fig. \ref{fig:laser}). The left setup locates inside of a shielding box where we perform DNP measurements on the diamond. The box shields from electrical and magnetic fields. The right setup locates outside of the box and includes the laser.
		
		The isolator is used to protect from the reflected light and to increase the beam size to a diameter of 5 mm to match the diamond size. The long pass filter is used to exclude lights except the red light emitting by the diamond. The lens, with focal length of 25.4 mm, is used to focus the emitted fluorescence, from diamond, to the photodetector’s sensor with a diameter of about 2 mm.
		
		\section{Experimental procedure}

		Before each DNP experiment, ODMR measurement was performed to find an exact positions of m$_s$ = $\pm$1 substates at a specific magnetic field. The diamond was irradiated with a wide range of MW sweep (up to 1.5 GHz) and with 532 nm laser light. The emitted red light from a diamond then was captured by a photo-detector. A data from the detector could be seen on a computer using a Labview program, where we could define exact MW frequencies corresponding to the m$_s$ = $\pm$1 substates and estimate a magnetic field.
		In this work we have only showed DNP measurements performed on NV centers aligned with the direction of the magnetic field. It has been problematic to align the diamond precisely, so that all 3 non-aligned NV orientations would have the same angle of 109.5$^o$ away from the direction of the magnetic field. So, we plan to improve our setup later and perform DNP through the non-aligned NV orientations, which will help us to increase the maximum hyperpolarization value by three times.
		All DNP experiments in this work have been performed at the magnetic field of about 17.6 mT, generated by the electrical magnet, which corresponds to a MW frequency of 2.370$\pm$0.005GHz for m$_s$ = -1 substate and 3.357$\pm$0.005GHz for m$_s$ = +1 substate of NV centers aligned with the magnetic field. $^{13}$C nuclear spins in the diamond was hyperpolarized for 100+ sec at constant MW irradiation and at constant laser irradiation with wavelength of 532 nm. Once $^{13}$C nuclei were hyperpolarized, the diamond was shuttled 80 cm up within 1 sec, by a high precision motion controller Newport XPS-RL, to the superconducting magnet, where the polarization was detected by the saddle coil. Then the signal, from the coil, was sent by LapNMR at the resonance frequency of 64.237MHz. The NMR measurement took 5 ms and then diamond was sent back to the previous position, within 2 sec, for additional polarization measurements. For a better SNR, the hyperpolarized $^{13}$C NMR signal was always averaged 10 times.
	
  \bibliography{apssamp}

	\end{document}